\documentstyle[aps,prl,psfig]{revtex}

\newcommand{\ra}{$1/f^\alpha$ noise }

\begin{document}
\draft
\twocolumn[\hsize\textwidth\columnwidth\hsize\csname @twocolumnfalse\endcsname
\title{A simple model for $1/f^\alpha$ noise}
\author{J\"orn Davidsen \cite{byline} and Heinz Georg Schuster}
\address{Institut f\"ur Theoretische Physik und Astrophysik, Christian-Albrechts-Universit\"at,\\
Olshausenstra\ss e 40, 24118 Kiel, Germany}
\date{\today}
\maketitle

\begin{abstract}

We present a simple stochastic mechanism which generates pulse trains exhibiting a power law distribution of the pulse intervals and a $1/f^\alpha$ power spectrum over several decades at low frequencies with $\alpha$ close to one. The essential ingredient of our model is a fluctuating threshold which performs a Brownian motion. Whenever an increasing potential $V(t)$ hits the threshold, $V(t)$ is reset to the origin and a pulse is emitted. We show that if $V(t)$ increases linearly in time, the pulse intervals can be approximated by a random walk with multiplicative noise. Our model agrees with recent experiments in neurobiology and explains the high interpulse interval variability and the occurrence of $1/f^\alpha$ noise observed in cortical neurons and earthquake data.

\end{abstract}
]

\narrowtext

The omnipresence of \ra in nature is one of the oldest puzzles in contemporary physics still lacking a generally accepted explanation. The phenomenon is characterized by a certain behavior of the respective time signal: The power spectrum $S(f)$ is proportional to $1/f^\alpha$ at low frequencies $f$ with $\alpha \approx 1$. Examples include the light of quasars \cite{pres}, electrical measurements \cite{kogan}, music and speech \cite{vos}, human cognition \cite{gil} and coordination \cite{yosh}, the current through ion channels \cite{bez}, network traffic \cite{uli}, burst errors in communication systems \cite{berger}, freeway traffic \cite{musha}, granular flow \cite{naka}, \emph{etc}.
The time signals of a large number of these systems \cite{yosh,uli,berger,musha,naka} resemble a pulse train consisting of individual, largely identical events which occur at discrete times.
This is especially true for spike trains of single nerve cells for which \ra has been observed in various brain structures \cite{tei89,tei96,yam86,gru,kod}. The reported exponents, which depend both on the presence or absence of a sensory stimulus \cite{tei89,tei96} and on the state of the animal (REM sleep vs awake state) \cite{yam86,gru,kod}, vary from 0.68 to 1.38. The power-law behavior for spike train power spectra lies within the range 0.01 to 10 Hz, extending typically over 2 decades. In almost all cases the upper limit of the observed time over which fractal correlations exist is imposed by the duration of the recording.

\begin{figure}[bt]
\centerline{\psfig{figure=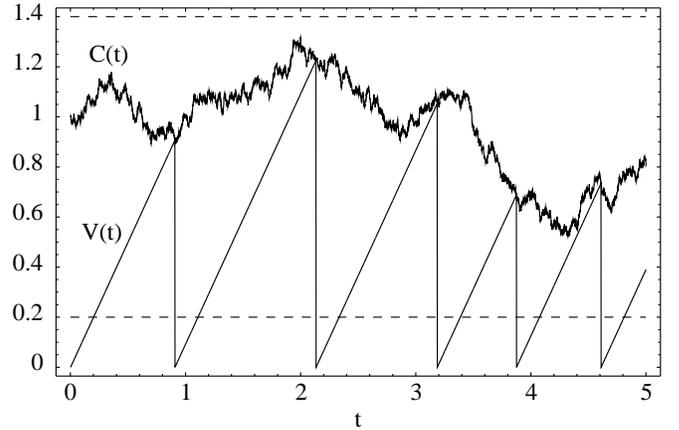,width=\columnwidth,clip=}}
\vspace{0.2cm}
 \caption{Dynamics of our model for a linear increasing voltage $V(t)$ with $V_0=0$. The dashed lines represent the lower and the upper boundary for the fluctuating threshold $C(t)$.}
 \label{picm}
 \end{figure}

In this Letter, we propose a simple mechanism for generating pulse trains with $1/f^\alpha$ behavior in systems with a threshold-controlled dynamics like, e.g., neurons and earthquake faults. Our model is based on an integrate-and-fire (IaF) mechanism and consists of a single unit characterized by two variables (see Fig. \ref{picm}): The voltage $V(t)$ and the threshold $C(t)$. Initially, the voltage is below the threshold. Then, the voltage increases monotonically in time --- in the simplest case just linearly --- and the threshold evolves according to a Brownian motion with diffusion constant $D$ within reflecting boundaries $V_0 < C_l < C(t) < C_u$. As soon as $V(t)$ has reached the threshold, the voltage is reset to $V_0$ and a pulse of unit height is emitted. In this way, a pulse train is generated. Note that the threshold is \emph{not} reset to its initial value.

Such a model is often used to describe single neurons: $V(t)$ is the membrane potential and the emitted pulse is the generated action potential. However, it is usually assumed that the threshold is constant in time. Recent investigation have shown that this is not true for cortical neurons \emph{in vivo} \cite{azo} and \emph{in vitro} \cite{hec}. Additionally, there is evidence that the spike trains of auditory neurons \cite{tei94} and of neurons in the mesencephalic reticular formation \cite{yam83} are not renewal, i.e., successive time intervals between spikes are correlated. These facts are incorporated in our model in the simplest possible way. Moreover, the effects of dead time or absolute refractoriness, which limits the rate at which a neuron can fire, are automatically included in our model via the lower bound $C_l$. The upper bound $C_u$ prevents an infinite time difference between two pulses.

\begin{figure}[bt]
\centerline{\psfig{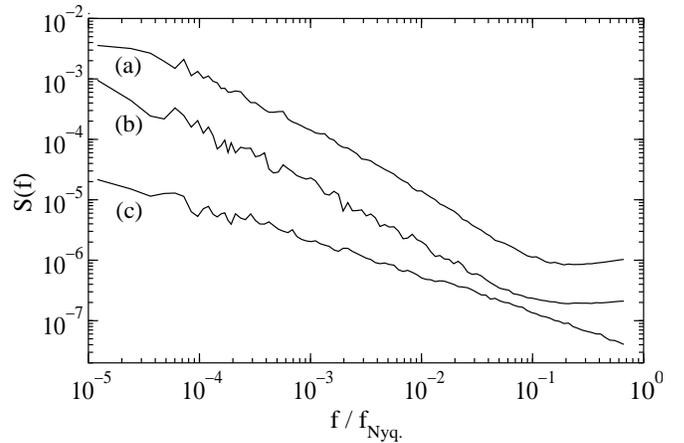}}
\vspace{0.2cm}
 \caption{Power spectrum of the pulse train generated by our model. Parameters are (a): $V_0 = 0$, $C_l = 0.2$, $C_u = 40$, $\protect\sqrt{D} = 0.2$. (b) as (a) with $C_u = 200$. (c): $V_0 = 0$, $C_l = 0.02$, $C_u = 4000$, $\protect\sqrt{D} = 2.0$. This curve is shifted down by 1 decade. All curves show a clear $1/f^\alpha$ decay. $\alpha$ varies from $0.6$ to $1.1$.}
 \label{pic_spec_all}
 \end{figure}

As one can see from Fig. \ref{pic_spec_all}, the power spectrum of the pulse train generated by our model with linear increasing voltage shows a $1/f^\alpha$ decay over several decades \cite{lin_text}. 
The frequency below which white noise behavior is observed is determined by $C_u - C_l$ and goes to zero for $C_u - C_l \to \infty$.
The exponent $\alpha$ is not universal and increases if the ratio $(C_l - V_0) / \sqrt{D}$ increases. In the limit $(C_l - V_0) \ll \sqrt{D}$, we find $\alpha = 0.5$. This result is explained by the fact that this limit corresponds to the case in which the waiting times between pulses are purely determined by the diffusive dynamics of the threshold. Hence, each waiting time is merely the first return time of a Brownian motion which obeys a power-law distribution with exponent $-1.5$. This implies $\alpha = 0.5$ \cite{schuster}.

To compare our results with real neurons, consider the parameters of curve (b): The dead time of a neuron is typically of order of milliseconds and the maximal time difference between two spikes of the order of seconds. This is exactly the ratio between $C_l$ and $C_u$. Moreover, we can now identify one unit of time in our model with 5 milliseconds real time. Hence, the $1/f^\alpha$ behavior in (b) is found for $f <$ 11 Hz. This and the fact that $\alpha \simeq 1.02$ reproduces the experimental results very well.

Extensive numerical investigations have shown that our model is very stable with respect to variations of the dynamics. Different forms of voltage increase, e.g., a linear, a squared, or a square-root increase, give similar results. The assumption of a monotonical increase of the voltage can also be dropped. A small amount of noise can be added to the voltage signal without altering our findings. Substituting the reflecting boundaries by a confining potential does not change our results, either. This points towards a generic behavior.

To obtain the power spectrum of our model analytically, consider the signal $X(t) = \sum_{k} \delta(t-t_k)$ where $t_k$ denotes the time of occurrence of the $k$th pulse. It follows as shown in \cite{dav00}
\begin{eqnarray}
S(f) &=& \lim_{T\to\infty} \frac{1}{2T} \left | \int\limits_{-T}^{T} dt X(t) \exp^{-i2\pi f t} \right |^2, \\
&=& \lim_{T \rightarrow \infty} \frac{1}{2T}  \sum_k \sum_q \exp^{i2\pi f (t_{k+q}-t_k)}.
\end{eqnarray}
With $\bar{I} = \lim_{T \rightarrow \infty} \frac{1}{2T} (k_{\rm max} - k_{\rm min} + 1)$, this leads to
\begin{eqnarray}
S(f) &\approx& \bar{I} \sum_q \left\langle \exp^{i2\pi f (t_{k+q}-t_k)} \right\rangle,
\label{equkm}
\end{eqnarray}
where $\langle \cdots \rangle$ denotes the average over the ensemble and over $k$. Hence, we need to know the probability distributions $\Psi_q(\tau) d\tau$ of the time differences between pulses $\tau = t_k - t_{k+q}$ for all integers $q$. $\Psi_q$ is merely the $q$th passage times density function $g_q$ averaged over the stationary probability distribution of ``initial'' states $V(0)$
\begin{eqnarray}
\Psi_q(\tau) = \langle g_q(\tau | V(0)) \rangle_{V(0)}.
\end{eqnarray}
Since $g_q$ can be computed from the first passage times density function (FPTDF), we will first focus on the latter.

The FPDTF of our model with \emph{linear} voltage increase can be obtained by mapping the model to an IaF model with constant threshold $C = (C_u + C_l) / 2$: The voltage $\tilde{V}(t)$ is defined as the sum of $V(t)$ and $C - C(t)$. This means that $\tilde{V}(t)$ fluctuates around $V(t)$. It also implies that $\tilde{V}(t)$ is reset to $V_0 + C - C(t_k)$ after the $k$th threshold crossing. Hence, the correlations are now encoded in the fluctuating reset. In conclusion, $\tilde{V}(t)$ behaves almost as a Brownian motion with drift to an absorbing barrier with a reset following each barrier crossing. The only difference is that the stochastic process is restricted to the interval $[V(t) + C - C_u, V(t) + C - C_l]$ at time $t$ which makes an exact mathematical treatment difficult. Neglecting the restriction for a moment and setting $C=0$, we obtain from the associated Fokker-Planck equation the FPTDF 
\begin{eqnarray}
g_1(\tau |\tilde{V}(0)) = \frac{-\tilde{V}(0)}{\sqrt{2 \pi D} \tau^{3/2}} \exp{-\frac{(\tilde{V}(0)+\tau)^2}{2 D \tau}},
\label{equ1}
\end{eqnarray}
and $g_1 \equiv 0$ for $\tau<0$. Here, $\tilde{V}(0)$ is the initial distance. Eq. (\ref{equ1}) is, of course, just an approximation to the FPTDF of our model. This can already be seen from the fact that the FPTDF of our model is identically zero for $\tau < (C_l - V_0)$ and $\tau > (C_u - V_0)$. However, as will be shown below this approximation proves to be very useful and Eq. (\ref{equ1}) can even be simplified to a Gaussian with the same mean and variance:
\begin{eqnarray}
g_1(\tau |\tilde{V}(0)) \propto \exp{-\frac{(\tilde{V}(0)+\tau)^2}{2 D |\tilde{V}(0)|}}.
\label{equ2}
\end{eqnarray}
From the FPTDF one could now compute higher passage times density functions $g_n$ in principle by convolution
\begin{eqnarray}
g_n(\tau |\tilde{V}(0)) = \int^\tau_0 g_1(t |\tilde{V}(0)) g_{n-1}(\tau-t| t) dt.
\label{equ3}
\end{eqnarray}
However, this convolution resists an analytical treatment due to the Markovian character of our process and the form of Eqs. (\ref{equ2},\ref{equ3}). The former manifests itself in the fact that the first passage time directly gives the initial distance between the threshold and the voltage for the next passage problem. The reason for this is that the norm of the $k$th reset of the voltage $\tilde{V}$ equals the time difference between the $k$th and the $(k-1)$th pulse due to the linear increase of the voltage $\tilde{V}$ with slope 1. Hence the reset depends on the last passage time only and the whole stochastic point process is totally described by the FPTDF plus its Markovian property.

 \begin{figure}[bt]
\centerline{\psfig{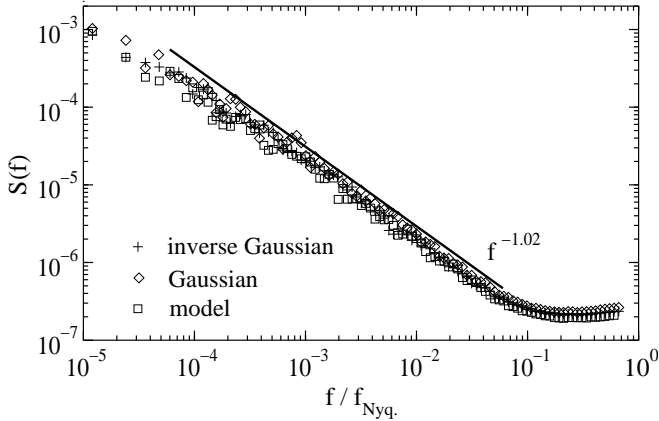}}
\vspace{0.2cm}
\caption{Power spectrum of the time signal generated by the linear version of our model and by the two approximations for the same parameters as in Fig. \ref{pic_spec_all}, (b).}
 \label{pic_spec_approx}
 \end{figure}

These properties enable us to interpret the stochastic process generated by the Gaussian approximation (\ref{equ2}) as a random walk of the inter-spike-intervals (ISI) $\tau_k$: 
\begin{eqnarray}
\tau_{k+1} = \tau_k + \sqrt{D \tau_k} \eta_k.
\label{rand}
\end{eqnarray}
Here, $\eta_k$ denotes the white noise source. Note the special kind of multiplicative noise which distinguishes our model from the one in \cite{kau}. The variance of the step length is proportional to the current ``position''. Hence, the origin is a fixed point of the random walk. To omit this difficulty and in spirit of our original model, we consider the random walk $\tau_k$ to be confined by two reflecting boundaries, i.e., $0 < (C_u - V_0) < \tau_k < (C_l - V_0)$. The power spectrum of the pulse-train generated by such a random walk with $t_{k+1} = t_{k} + \tau_k$ shows a clear $1/f^\alpha$ behavior with the same exponent as for our model (see Fig. \ref{pic_spec_approx}). This is also true if the new inter-spike-interval is chosen from an inverse Gaussian distribution given in Eq. (\ref{equ1}) with an initial condition depending on the last interval. Hence, these approximations seem to be justified. This is further confirmed by the stationary ISI distribution. For the random walk, the ISI distribution function $P(\tau)$ is proportional to $\tau^{-1}$ \cite{isi_text}. Simulations show that this is in excellent agreement with our model and the inverse Gaussian approximation.

We have to point out that the behavior of $P(\tau)$ depends on the specific kind of voltage increase. For a voltage increase with $(t-t_{last})^\beta$ with $0 < \beta < 2$, we find $P(\tau) \propto \tau^{\beta-2}$ for our model. Substituting the reflecting boundaries by a potential of the form $a/x + b x^2$ for example, we still find power law tails in the ISI distribution. This is exactly what measurements show for cortical neurons \cite{wise}, especially in the mesencephalic reticular formation \cite{yam83}. Hence, our model seems to be especially well-suited to explain the occurrence of \ra in that formation. In general, IaF models can not explain the high interspike interval variability exhibited by cortical neurons \cite{sof}. A fluctuating threshold as described by our model, however, can solve this long-standing issue.

For renewal processes the ISI distribution function and the FPTDF are identical and completely describe the process. This is the case for a standard IaF neuron with reset of the voltage to the origin after each barrier crossing which is described by the FPTDF in Eq. (\ref{equ1}). Such a model can neither explain a $1/f^\alpha$ signal nor a power-law decay of the ISI distribution. In contrast to our model, Usher and co-workers showed that fractal behavior might be a consequence of the global activity dynamics of a network of IaF neurons \cite{ush}. Due to experimental limitations, however, the link between $1/f^\alpha$ single-unit power spectra and macroscopic activity dynamics remains, as of now, a conjecture. Another attempt to explain the phenomenon of \ra is based on fractal and fractal-rate stochastic point processes \cite{low}. Certain types of these processes properly characterize the statistical properties found in different experiments. However, in many cases it is not clear \emph{a priori} why the real system should generate such a process. This is especially true for the clustering Poisson process which was applied to explain \ra in the mesencephalic reticular formation \cite{gru}. In contrast to that, our model provides a simple explanation of \ra in integrate-and-fire systems in general and, hence, applies to neurons in particular. The crucial assumption is a fluctuating threshold. Such a behavior is related to models presented in \cite{haen} and was already considered in \cite{wise} to explain the power law decay of neuronal ISI distribution functions. However, the stochastic point process was still assumed to be renewal and could not explain a $1/f^\alpha$ behavior of the power spectral density function. Consequently, the second crucial assumption of our model is the Markovian character of the process in agreement with Refs. \cite{dav00,kau}. To directly verify our model for neurons in the mesencephalic reticular formation, one should measure the evolution of the reset from the threshold potential.

\begin{figure}[bt]
\centerline{\psfig{figure=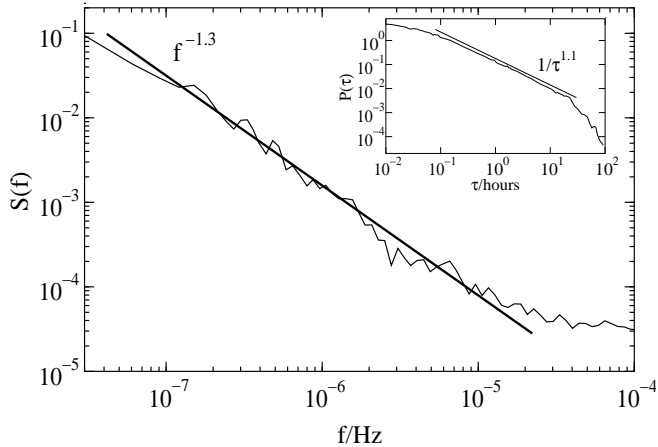,width=\columnwidth,clip=}}
\vspace{0.2cm}
 \caption{Power spectrum of the earthquake signal. Inset: ISI distribution.}
 \label{geo}
 \end{figure}

All three main characteristics of our model, i.e., a $1/f^\alpha$ spectrum, a power-law decay of the ISI distribution, and a Markovian dependence of the ISI, can not only be observed in nerve cells but also in other systems. For example, the pulse signal, defined by associating the occurrence of pulses of unit height with earthquakes in the Mojave region \cite{mojave}, shows $1/f^\alpha$ fluctuations with $\alpha=1.3$ and has an ISI distribution with exponent $-1.1$ (Fig. \ref{geo}). Moreover, there is evidence that the ISI do not obey a renewal process. Rather a Markovian dynamics can be found \cite{dav01}. These findings imply that the effective evolution of the ISI is similar to the one generated by our model and caricatured by Eq. (\ref{rand}).

To summarize, we have presented a simple stochastic model which is able to transform integrated white noise with a $1/f^2$ power spectrum into noise with $1/f$ tail. The basic mechanism is similar to what is expected to describe the dynamics of single neurons and, thus, our model can explain the behavior of neurons in the central-nervous-system \cite{gru,yam83}. 
The analysis of earthquake data even suggests that the effective dynamics of the ISI is the same for many systems exhibiting $1/f^\alpha$ noise.

We thank A. Aertsen and C. Goltz for useful discussions. J. D. would like to thank the Land Schleswig-Holstein, Germany, for financial support.


\begin{references}
\bibitem{lin_text}
The slope of the voltage increase is not relevant and is set to 1. We assume $C_l-V_0 \ll C_u-C_l$ and $0 \ll D \ll C_u-C_l$ to exclude trivial behavior.
\bibitem{isi_text}
The random walk with multiplicative noise can be approximated by a random walk with constant variance and non-uniform waiting times between steps which are proportional to $\tau_k^{-1}$. This explains the power-law decay of $P(\tau)$. Note that deviations from this behavior occur close to the boundaries.
\bibitem{mojave}
The investigated time signal contains all earthquakes above magnitude 1.8 in the area of $34^\circ 0'$ lat. N to $35^\circ 45'$ lat. N and $115^\circ 45'$ lon. W to $118^\circ 15'$ lon. W from 01/01/84 till 12/31/00 which can be found in the Southern California Seismographic Network Catalogues.
\end{references}
\end{document}